\documentclass{ws-p8-50x6-00}

\begin{document}

\title{Status of Hard Interactions \\ (Jets and Heavy Flavor)}

\author{M. Klasen}

\address{II.\ Institut f\"ur Theoretische Physik, Universit\"at Hamburg,
Luruper Chaussee 149, D-22761 Hamburg, Germany\\
E-mail: klasen@mail.desy.de}

\maketitle

\abstracts{
We review the status of hard interactions, in particular of jet and heavy
flavor production, at HERA and LEP. Emphasis is given to recent theoretical
developments. Instantons, event shapes, and prompt photons are also
briefly discussed.}

\section{Introduction}

\vspace*{-7.5cm}
DESY 01-089 \\
hep-ph/0106274
\vspace*{6.5cm}

At the DIS 2000 conference in Liverpool, a large variety of perturbative QCD
calculations for jet and heavy flavor production was confronted with
experimental data from HERA, LEP, and the TEVATRON. In many cases, though
not all, reasonable agreement between theory and experiment was found, thus
constituting a successful test of QCD as the theory of strong interactions.
The further reaching goals of extracting and testing the non-perturbative
fundamental parameters of this theory were, however, not reached at the same
level. Considerable room for improvement was left for the determination of the
strong coupling $\alpha_s(Q^2)$, the hadron and photon fragmentation functions,
and the structure functions of the proton and in particular of the photon.
Experimental results for the polarization of the $J/\Psi$ charmonium
shed new doubt on the theory of heavy quark-antiquark bound states,
Non-Relativistic QCD (NRQCD), and the precise values of its non-perturbative
operator matrix elements remained unclear. In many cases DIS 2000 had to leave
as desiderata the reduction of the experimental and theoretical uncertainties,
in particular in the hadronic energy scale and in the dependence on the
renormalization and factorization scales.~\cite{Blair:2000pt}
\begin{figure}[t]
 \begin{center}
  \epsfig{file=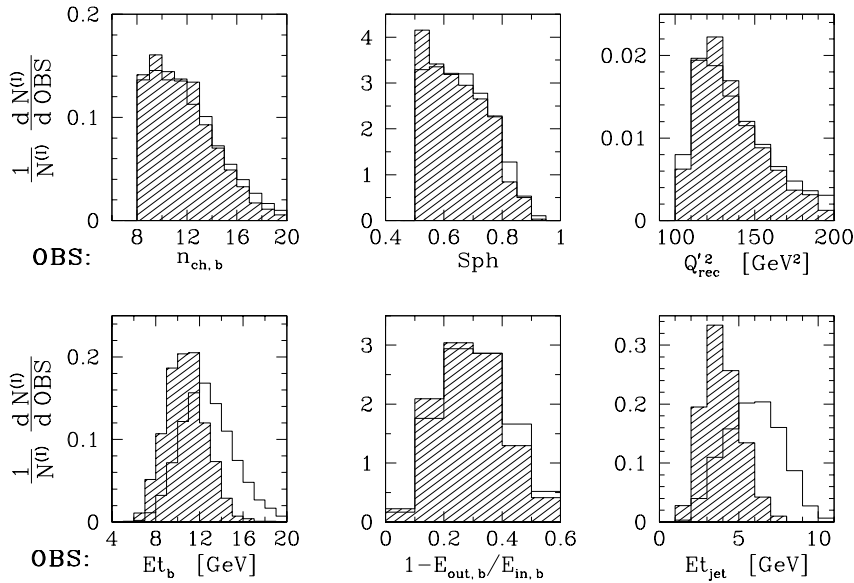,width=0.65\textwidth,angle=-90}
 \end{center}
\caption{\label{fig:1}Dependence of the shape normalized distributions of the
six H1 instanton observables on the low-$Q^2$ portion of the spectrum. The
shaded distributions are obtained with an additional cut $x\geq
0.15$.~\protect\cite{Ringwald:2001gt}}
\end{figure}

The experimental errors can often be reduced by considering rates instead of
absolute cross sections. Theoretical progress is harder to reach and requires
tedious next-to-leading order (NLO), next-to-next-to-leading order (NNLO), and
resummed calculations and may also necessitate the inclusion of power
corrections. Sufficient theoretical interest is thus mandatory for precise
determinations of parton densities etc., which also form the basis for reliable
estimates of signals and backgrounds of physics beyond the Standard Model.
In this paper we review progress along these lines at and prior to the DIS
2001 conference in Bologna. The paper is organized from the general to the
detailed features of the hadronic final state. In Sec.\ 2, the current status
of instantons and event shapes is discussed. Sec.\ 3 summarizes recent progress
for jet production and Sec.\ 4 for prompt photon production. Open and
bound-state heavy flavor production is discussed in Sec.\ 5, and Sec.\ 6
contains a brief summary.

\section{Instantons and Event Shapes}

At DIS 2000 a first dedicated search for instantons was presented by the
H1 collaboration. It yielded an excess of $549$ observed events over 
$435^{+36}_{-22}$ and $363^{+22}_{-26}$ events expected from matrix elements
plus parton showers (MEPS) and the color dipole model
(CDM).~\cite{Mikocki:2000xf} Four of the six observables agreed in shape with
the expected instanton signal, but the two distributions in the transverse
energies of the instanton band ($Et_b$) and of the current jet ($Et_{\rm jet}$)
deviated. The discrepancy could now be traced back to the fact that a fiducial
cut on $x\geq 0.15$, which depletes the theoretically unreliable low $Q^2$
region, was missing in the experimental analysis.~\cite{Ringwald:2001gt}
As can be seen in Fig.\ \ref{fig:1} this affects exactly the two transverse
energies $Et_b$ and $Et_{\rm jet}$, which should agree better in shape with
the data if the cut $x\geq 0.15$ is implemented.
\begin{figure}[t]
 \begin{center}
  \epsfig{file=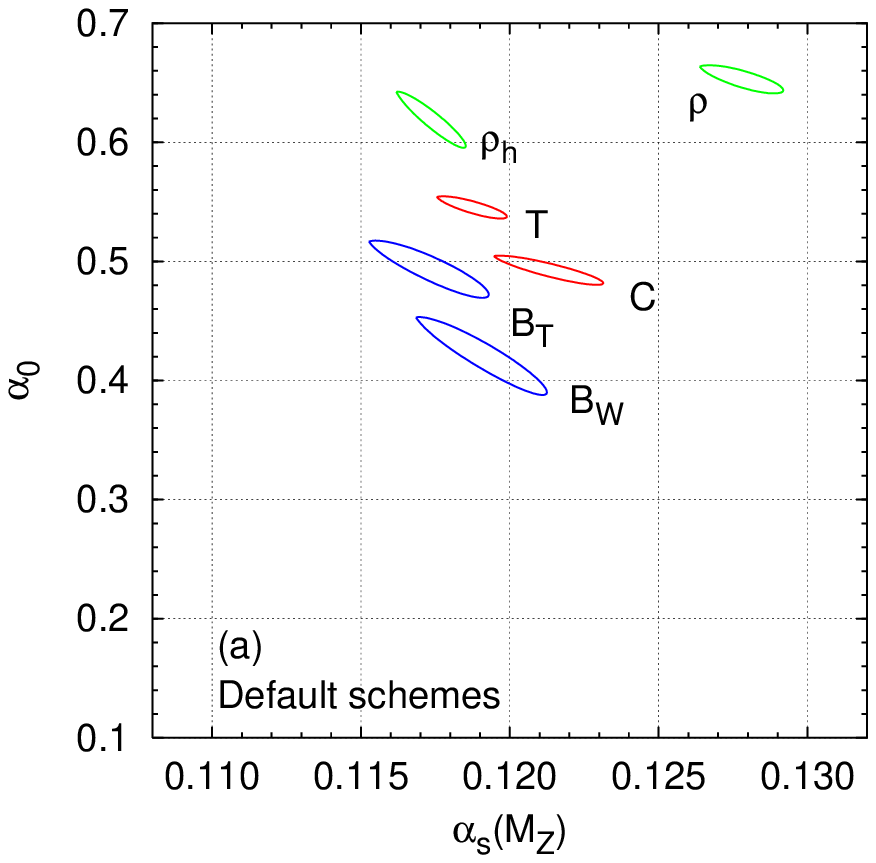,width=0.485\textwidth}%\;
  \epsfig{file=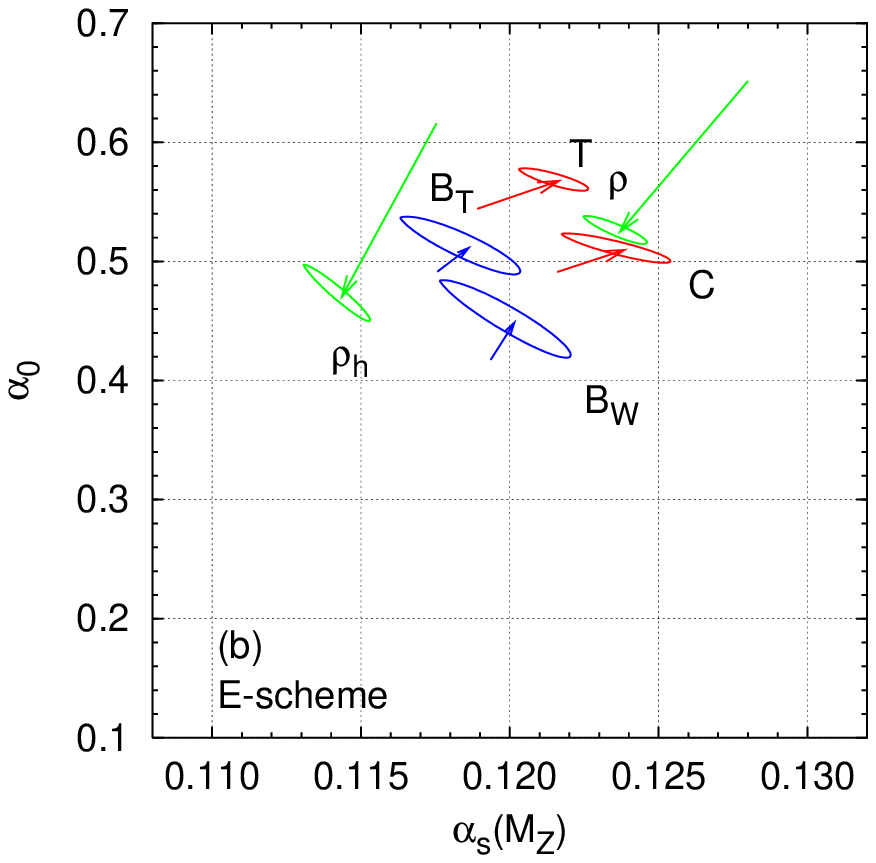,width=0.485\textwidth}
  \vspace{0mm} \\
  \epsfig{file=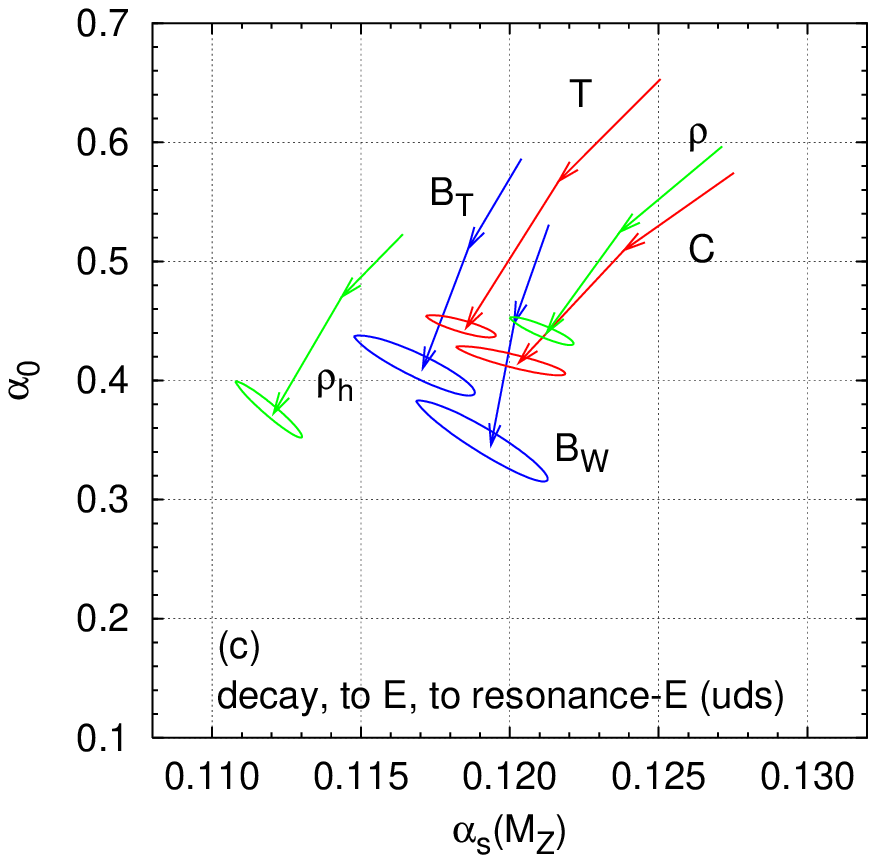,width=0.485\textwidth}%\;
  \epsfig{file=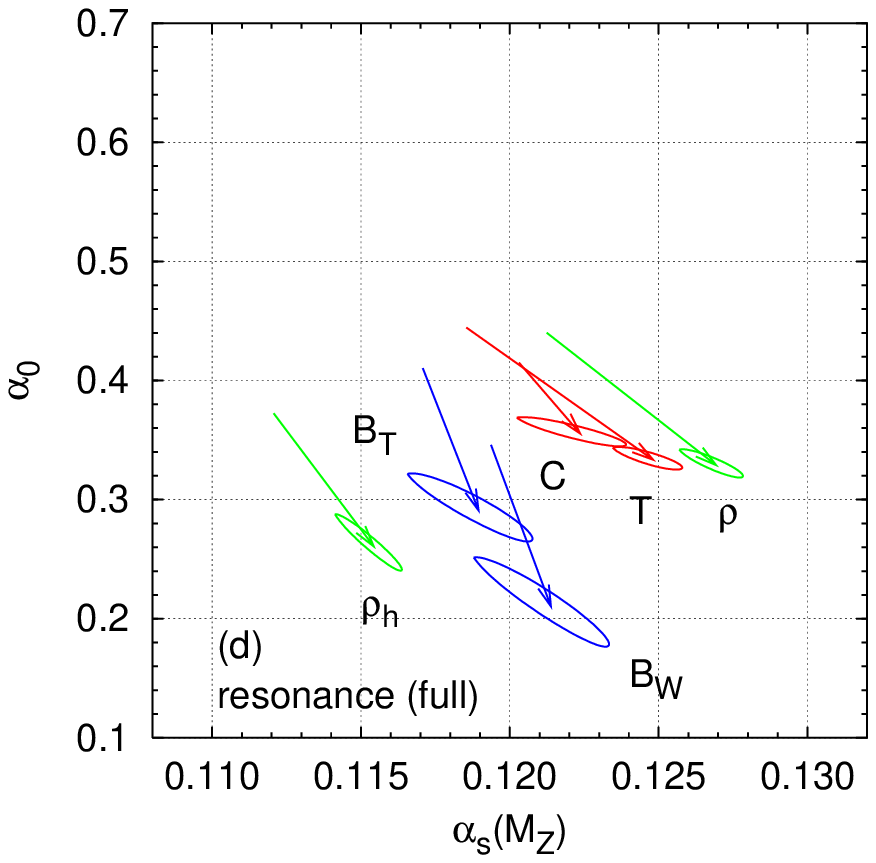,width=0.485\textwidth}
 \end{center}
\caption{\label{fig:2}One-$\sigma$ confidence level contours from fits to
event shape variables in the default (a) and $E$-scheme (b) at the normal
hadron level, and in the $E$-scheme corrected to the resonance level for
light (c) and light plus heavy (d) quarks.~\protect\cite{Salam:2001bd}}
\end{figure}
\begin{figure}
 \begin{center}
  \epsfig{file=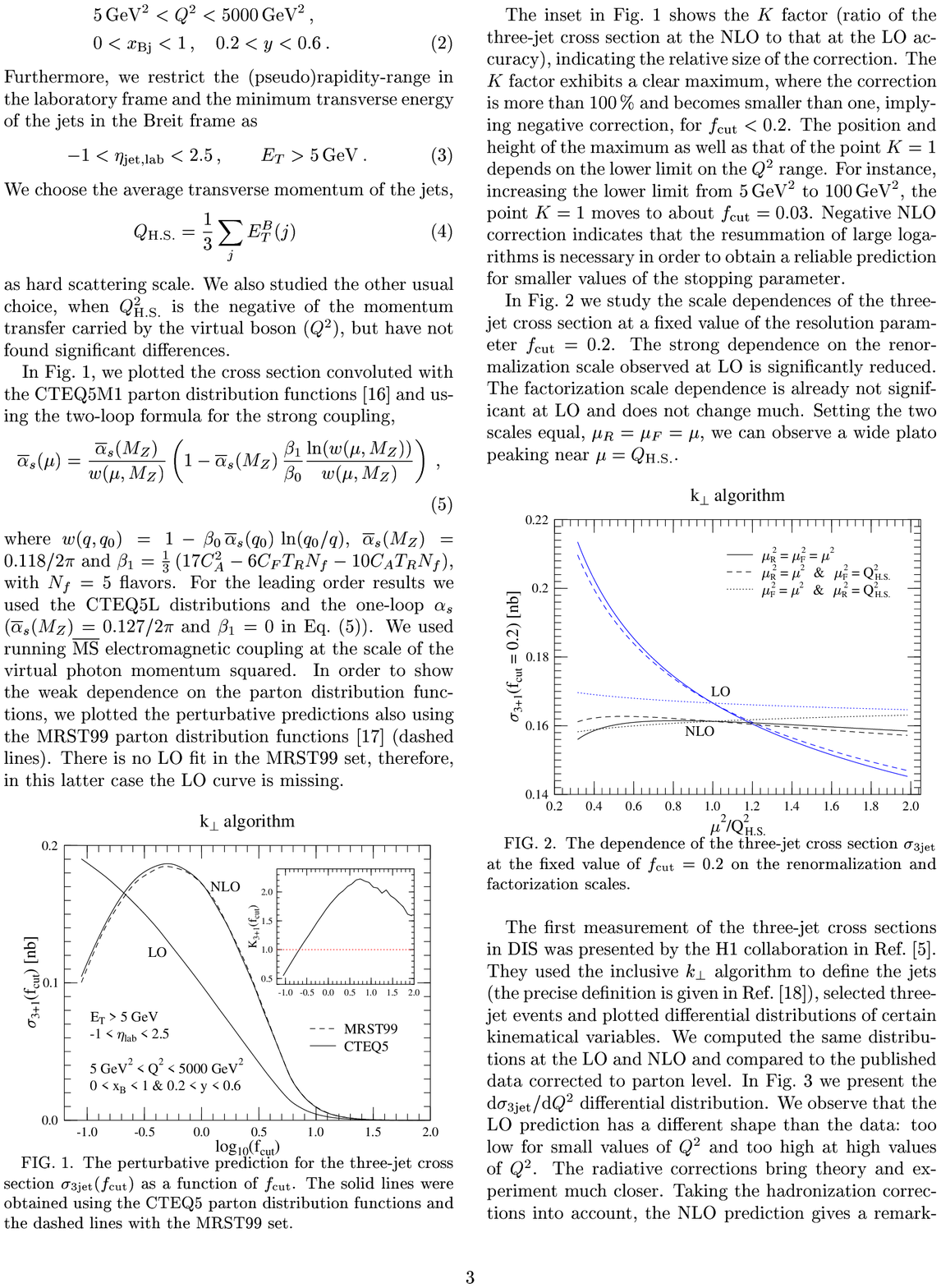,bbllx=320pt,bblly=337pt,bburx=555pt,bbury=528pt,%
          width=\textwidth,clip=}
 \end{center}
\caption{\label{fig:5}Scale dependence of the NLO three-jet cross section in
 DIS.~\protect\cite{Nagy:2001xb}}
\end{figure}
The event shape variables thrust $T$, jet mass $\rho$, jet broadening $B_T$
and $B_W$, and the $C$-parameter provide a means to describe general features
of a multi-particle hadronic final state in $e^+e^-$ annihilation.
The dominant contributions from low
multiplicities can be calculated perturbatively, but these calculations 
become sensitive to hadronization and power corrections at the edges of
phase space. A universal description of the event shape variables is possible
with the help of the running strong coupling $\alpha_s(Q^2)$ and its effective
value in the infrared regime $\alpha_0$. Recently the event shape variables
have been shown to depend on the particle masses and thus on their definitions
in terms of momenta ($p$-scheme) or energies ($E$-scheme) and also on the decay
level of the particles (resonance or decay).~\cite{Salam:2001bd}
In the $E$-scheme mass effects are absent and, as Fig.\ \ref{fig:2}
demonstrates, the uncertainties in the $\alpha_s(Q^2)$ and $\alpha_0$
determinations from different event shape variables are reduced.

\begin{figure}
 \begin{center}
  \epsfig{file=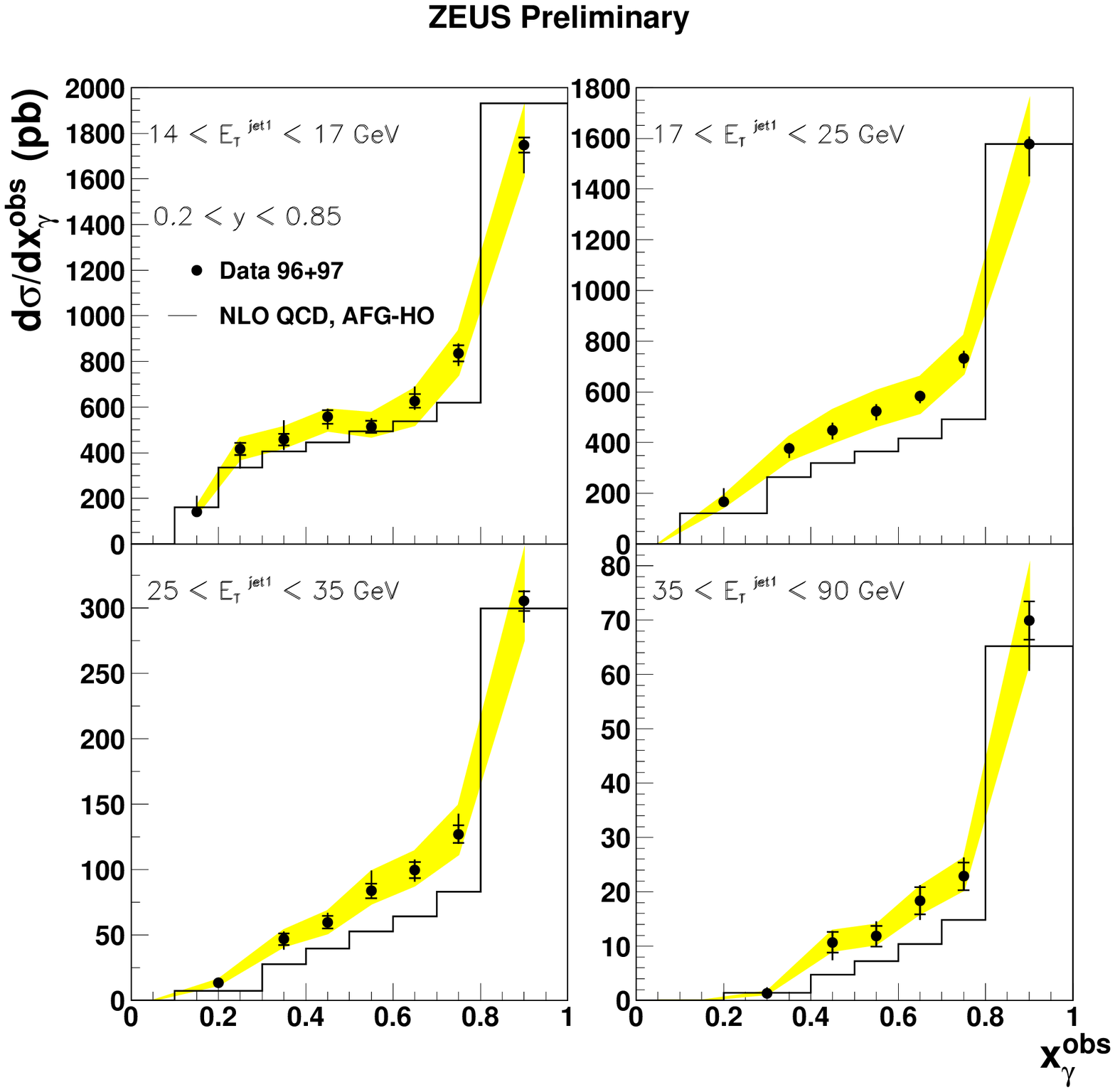,bbllx=0pt,bblly=0pt,bburx=561pt,bbury=551pt,%
          width=0.48\textwidth,clip=}
  \epsfig{file=opal.ps,bbllx=21pt,bblly=150pt,bburx=525pt,bbury=677pt,%
          width=0.45\textwidth,clip=}
 \end{center}
\caption{\label{fig:xgamma}The NLO dijet photoproduction cross section as a
function of the observed momentum fraction of the partons in the photon
compared to ZEUS and OPAL data.~\protect\cite{zeus1066,opal443}}
\end{figure}
Similar event shape variables can be defined in deep-inelastic scattering,
where non-global observables require special care in resummed calculations
due to soft emission between hemispheres.~\cite{Dasgupta:2001sh}
A discrepancy in the thrust distribution d$\sigma/$d$\tau_z$ between the
perturbative DISENT~\cite{Catani:1997vz} prediction on the one hand and the
DISASTER++~\cite{Graudenz:1997gv} and resummed
results on the other hand, which was uncovered at DIS 2000, is unfortunately
still unresolved.~\cite{Dasgupta:2000bs}

\section{Jet production in DIS, $\gamma p$, and $\gamma\gamma$}

The strong coupling $\alpha_s(M_Z)$ can be determined in DIS not only in
structure function measurements, but also in jet production. Recent
determinations by the H1 and ZEUS collaborations are in good agreement with
the world average and have quite competitive error bars.~\cite{grindhammer}
Both the experimental and theoretical errors in these determinations could
be reduced, since jet rates were analyzed instead of absolute jet cross
sections. Nevertheless, the uncertainty in the H1 three-jet
rate from the LO scale variation was still larger than the sensitivity to
$\alpha_s(M_Z)$. This could now be changed since a NLO three-jet calculation
in DIS has recently become available.~\cite{Nagy:2001xb} The calculation uses
helicity amplitudes, crossed from the result $e^+e^-\rightarrow 4$ jets, which
have been calculated by several groups, and it has been numerically checked
against the NLO dijet result of DISASTER++.~\cite{Graudenz:1997gv}
How much the NLO corrections reduce the scale dependence can clearly be seen
in Fig.\ \ref{fig:5}.

\begin{figure}
 \begin{center}
  \epsfig{file=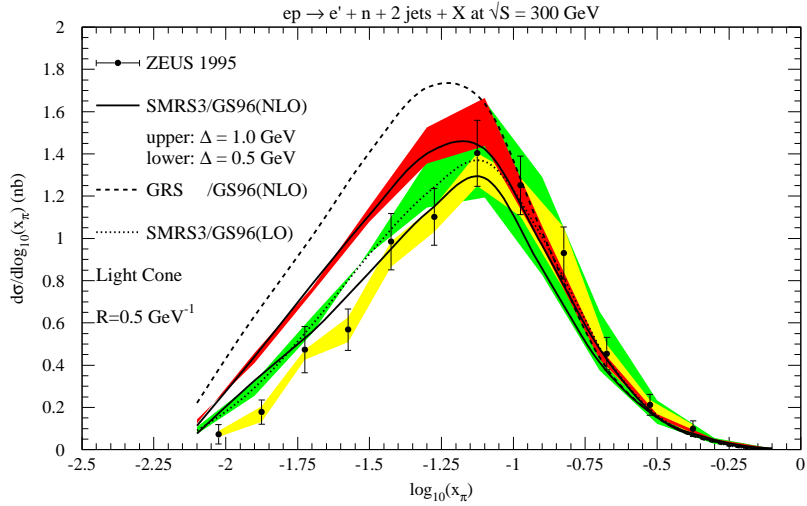,width=\textwidth}
 \end{center}
 \caption{\label{fig:xpi}
 The NLO dijet photoproduction cross section with a leading neutron as a
 function of the observed momentum fraction of the partons in the pion compared
 to ZEUS data.~\protect\cite{Klasen:2001sg}}
\end{figure}
For jet production in photon-proton collisions three different NLO calculations
have become available over the last
years.~\cite{Klasen:1997it,Harris:1997hz,Frixione:1997ks} Although they use
different
phase-space slicing and subtraction techniques, they agree well within the
numerical accuracy.~\cite{Harris:1998ss} Now a fourth NLO dijet calculation
has become available which uses a phase space slicing technique in transverse
energy $E_T$ and jet radius $R$.~\cite{Aurenche:2000nc} This calculation
has also been found to agree with the calculation by Klasen and
Kramer.~\cite{Klasen:1997it}
Furthermore it has confirmed previous observations that dijet cross sections
defined with an identical cut on both jet transverse energies are infrared
sensitive due to incomplete cancellations of virtual and real
corrections.~\cite{Klasen:1996xe} In addition, Aurenche {\it et al.} remark
that this sensitivity propagates into the variable $x_\gamma^{\rm obs}=
\sum_{i=1,2} E_{T,i}e^{-\eta_i}/(2E_\gamma)$ and propose to replace the 
transverse energy of the second jet by that of the first jet. However,
this prescription still does not cure the (integrable) singularity at
$x_\gamma^{\rm obs}=1$ so that it remains necessary to choose a large bin
width in $x_\gamma^{\rm obs}$. This is demonstrated in Fig.\ \ref{fig:xgamma},
where recent ZEUS data \cite{zeus1066} with a bin width of 0.2 and OPAL
data \cite{opal443} with a bin width of 0.1 are compared to NLO QCD
predictions using AFG and GRV photon densities. At small $x_\gamma^{\rm obs}$,
both experiments find lower cross sections than suggested by the used photon
densities.

\begin{figure}
 \begin{center}
  \epsfig{file=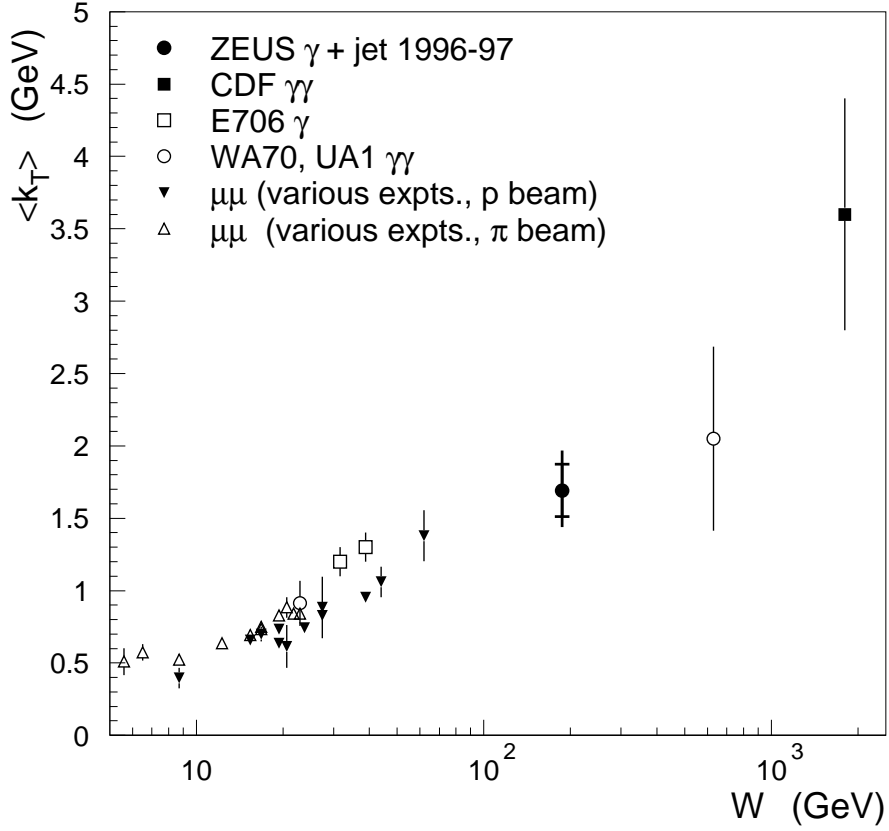,width=\textwidth,clip=}
 \end{center}
\caption{\label{fig:prompty}Intrinsic $k_T$ in prompt photon production
 at various hadronic energies $W$.~\protect\cite{Chekanov:2001aq}}
\end{figure}
At HERA jets can be produced in photoproduction in association with a leading
neutron. If the energy transfer between the proton and neutron is small,
the cross section is dominated by virtual pion exchange and the dijet
cross section can be used to determine the parton densities in the
pion.~\cite{Klasen:2001sg} In Fig.\ \ref{fig:xpi} the NLO dijet photoproduction
cross section with a leading neutron is shown as a function of the observed
momentum fraction of the partons in the pion. Unfortunately, equal cuts on
$E_T$ have been used in the ZEUS measurement~\cite{Breitweg:2000nk}, but the
SMRS3~\cite{Sutton:1992ay} pion densities seem to agree better with the
data than the GRS~\cite{Gluck:1999xe} densities.

\begin{figure}
 \begin{center}
  \epsfig{file=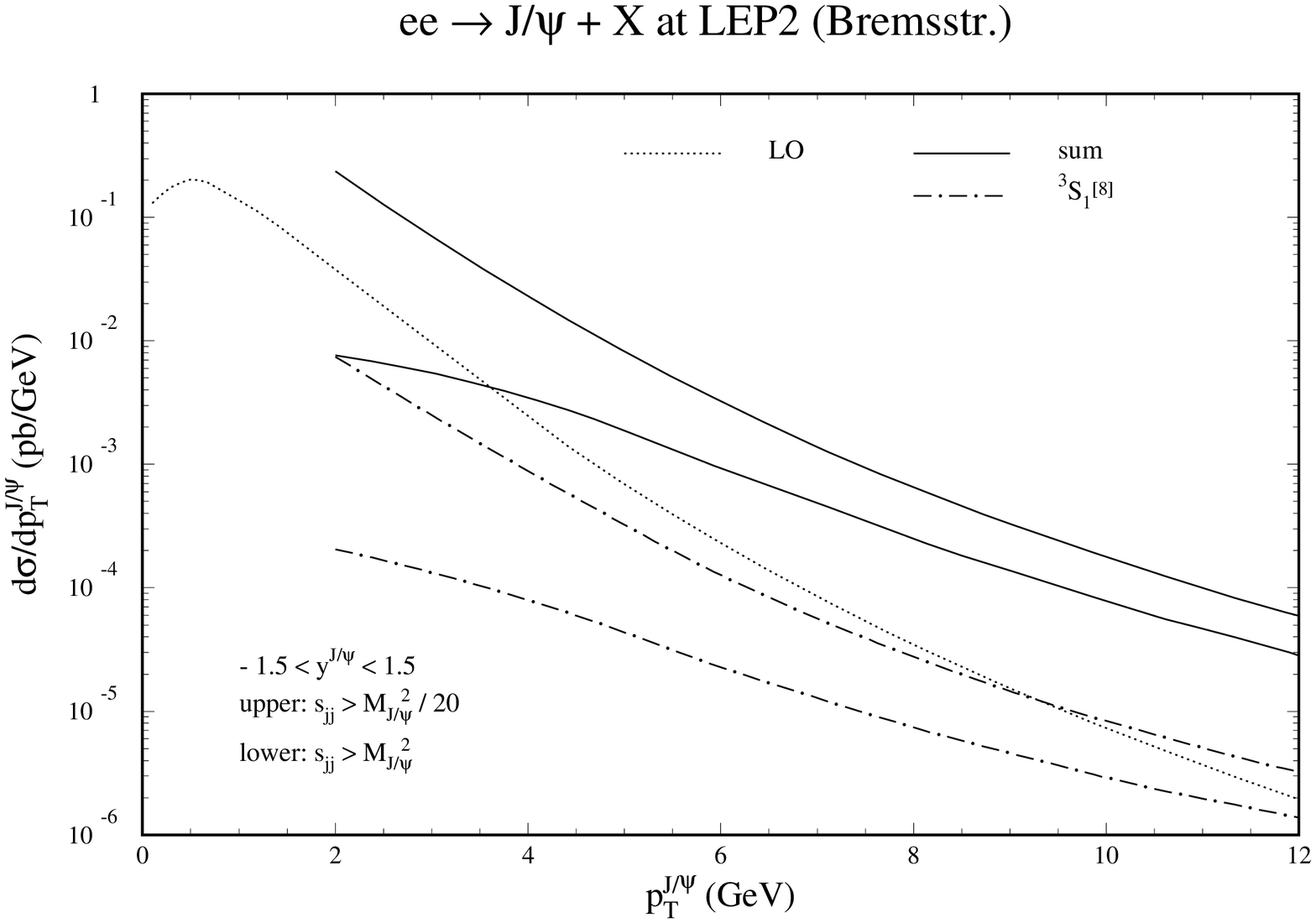,width=\textwidth,clip=}
 \end{center}
\caption{\label{fig:jpsinlo}Direct $J/\Psi$ production in photon-photon
collisions as a function of $E_T$. The real corrections are large at large
$E_T$, where the uncertainty from the invariant mass regulator is 
reduced.~\protect\cite{Klasen:2001mi}}
\end{figure}
\section{Prompt photons in photoproduction}

Prompt photon production is known to suffer from uncertainties
due to fragmentation contributions, isolation ambiguities,
and possibly intrinsic transverse momenta of the scattering initial partons.
ZEUS have analyzed prompt photon photoproduction and observed an excess over
the LO QCD expectation in the $p_\bot,p_\|$, and $\Delta\phi$
distributions.~\cite{Chekanov:2001aq}
 They have then attributed these effects to an effective $\langle k_T
 \rangle$ of 1.69 $\pm 0.18^{+0.18}_{-0.20}$ GeV, which includes effects
 coming from the initial-state parton showering as modelled within PYTHIA.
 This value of $\langle k_T
 \rangle$ seems to be consistent with determinations in hadron collisions
 at different energies (see Fig.\ \ref{fig:prompty}), which are, however,
 obtained using a variety of methods. The data could not yet be confronted
 with a complete NLO QCD calculation \cite{Fontannaz:2001ek} or resummed
 QCD results. Such a comparison will be complicated by the fact that equal
 cuts on the transverse energies of the photon and recoiling jet have been
 used ($E_T>5$ GeV). Nevertheless, the ZEUS data seem to support the trend
 that the effective $\langle k_T \rangle$ in the proton rises with the
 available hadronic energy.

\begin{figure}
 \begin{center}
  \epsfig{file=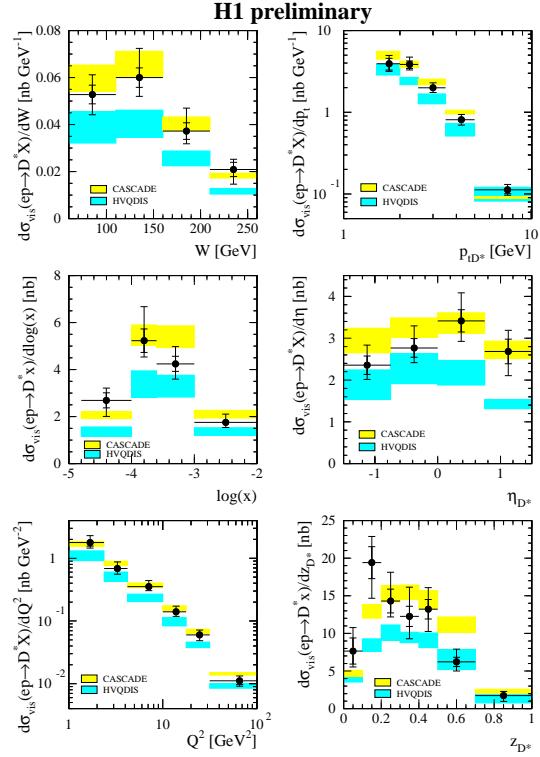,bbllx=75pt,bblly=143pt,bburx=455pt,bbury=696pt,%
          width=0.59\textwidth,clip=}
 \end{center}
\caption{\label{fig:h1dstar}Preliminary H1 data on $D^\ast$ production in DIS,
compared to the NLO QCD calculation HVQDIS and the CASCADE Monte Carlo.~\protect\cite{h1984}}
\end{figure}
\section{Heavy flavor production}

The production of heavy quark-antiquark bound states is described by
the effective field theory of Non-Relativistic QCD (NRQCD). It predicts
the production of intermediate color-octet bound states, which transform
non-perturbatively into the observed color-singlet quarkonia by soft
gluon radiation. The color-octet mechanism is clearly needed to describe
the transverse momentum spectra of $J/\Psi$ and $\Psi'$ mesons at the 
TEVATRON. However, recent ZEUS data confirm that it is not needed in
photoproduction, where the NLO color-singlet prediction is fully sufficient
to describe the data.~\cite{zeus851}
To clarify the situation NLO corrections have to be included in other
quarkonium processes like photon-photon collisions. In this case direct,
single-resolved, and double-resolved processes contribute at large,
intermediate, and small $p_T$.~\cite{Godbole:2001pj} The real corrections
to the direct channel have recently been calculated \cite{Klasen:2001mi} and
are shown in Fig.\ \ref{fig:jpsinlo}. At large $p_T$ they increase the direct
LO cross section by about an order of magnitude.

Open charm production has been measured in DIS by the ZEUS and H1 
collaborations. While ZEUS find good agreement in the $Q^2$ and $x$
distributions \cite{zeus855}, H1 see an excess over NLO QCD predictions
\cite{Harris:1995tu}
using GRV parton densities and the Peterson fragmentation function
at small $p_T$ and forward rapidities $\eta$ (see Fig.\
\ref{fig:h1dstar}).~\cite{h1984}
Unfortunately the data are not sensitive enough to distinguish between
different fixed
and variable flavor schemes.~\cite{Chuvakin:2001zj}

\begin{figure}
 \begin{center}
  \epsfig{file=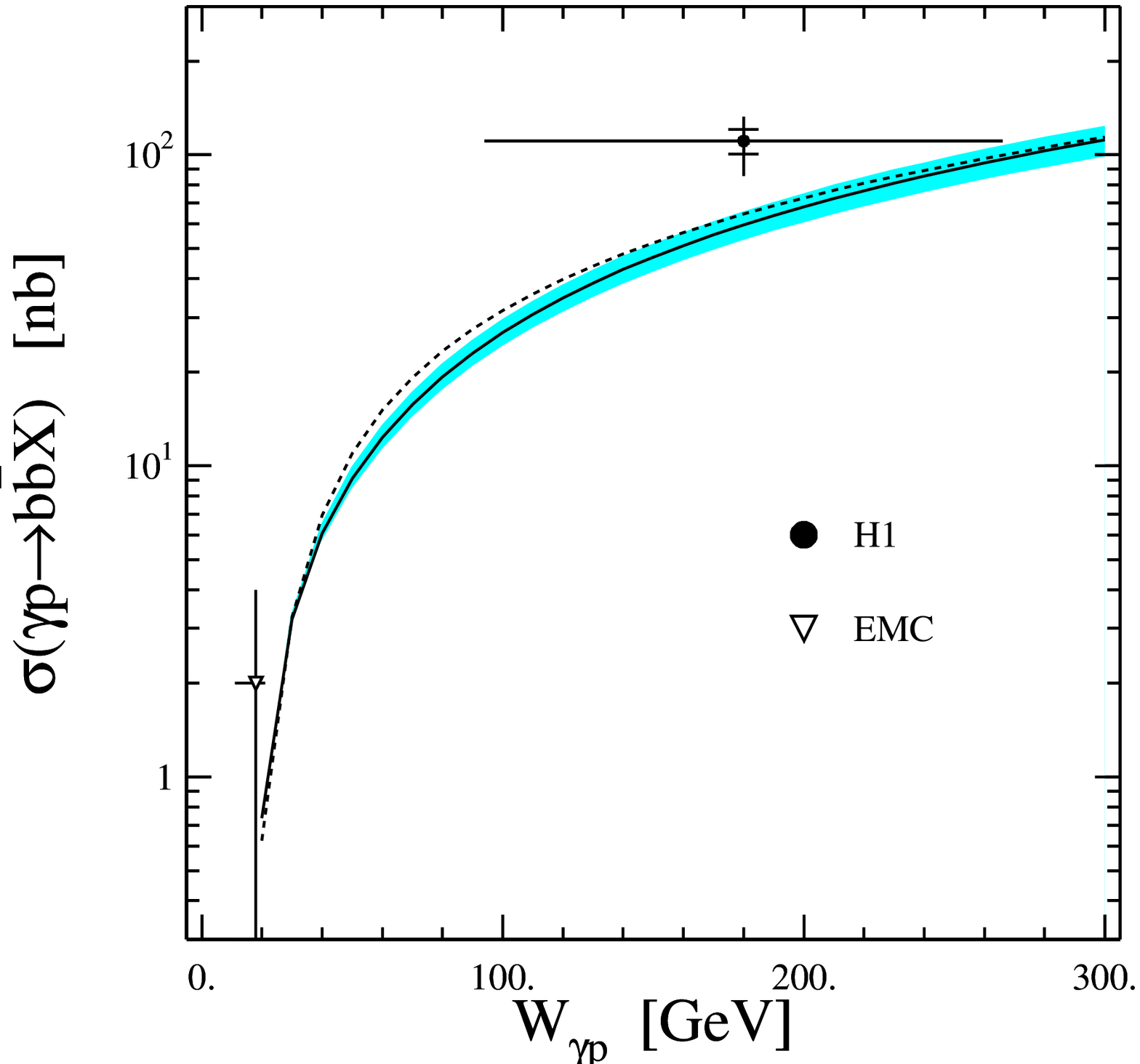,width=0.38\textwidth,clip=}
  \epsfig{file=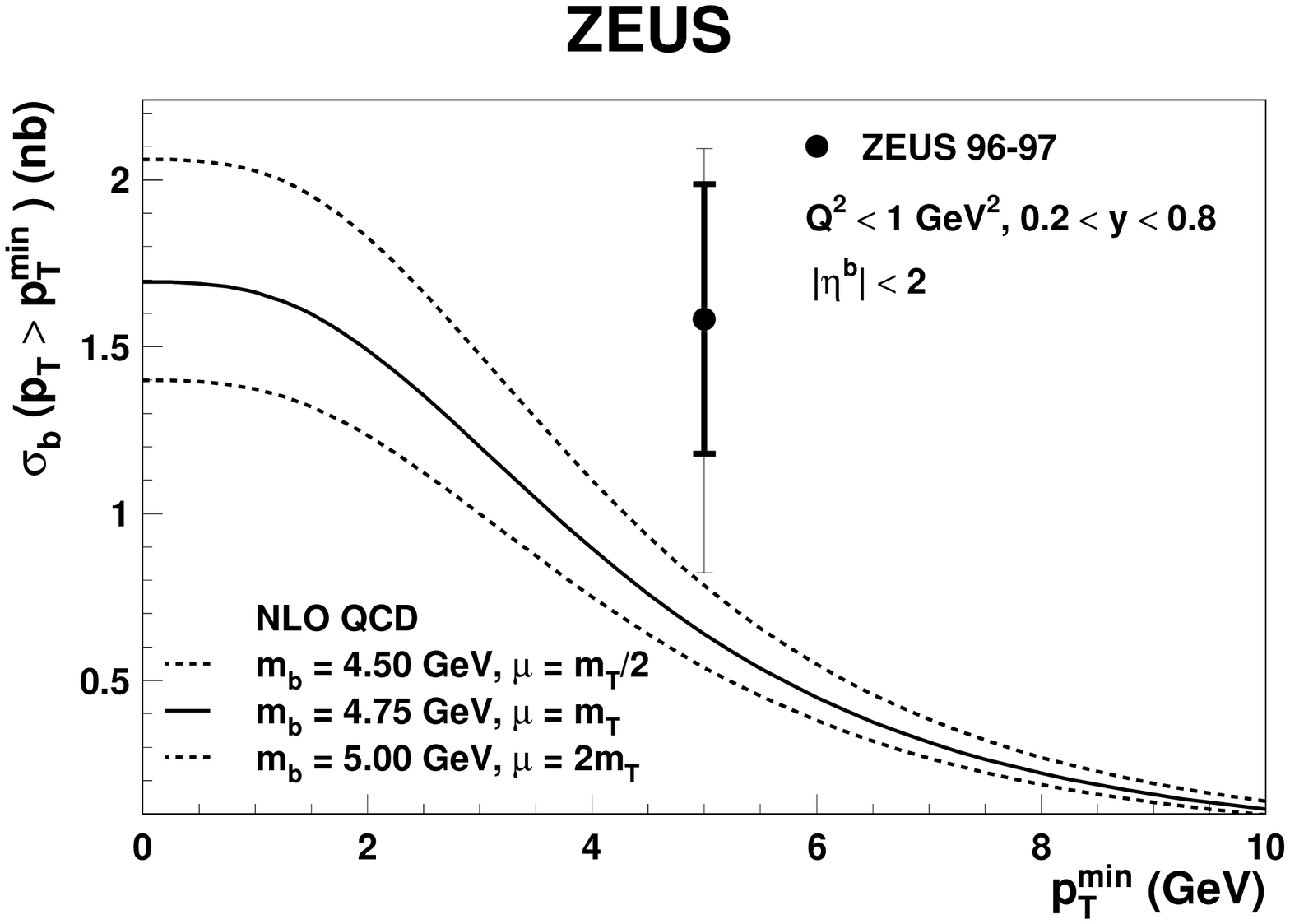,width=0.51\textwidth,clip=}
 \end{center}
\caption{\label{fig:bottom}H1 and ZEUS measurements of the total bottom
cross section in photoproduction.~\protect\cite{Adloff:1999nr,Breitweg:2001nz}}
\end{figure}

A particularly interesting topic is the bottom cross section, which is 
in excess over massive QCD predictions in hadron-hadron, photon-hadron,
and photon-photon collisions (see Fig.\ \ref{fig:bottom} for the HERA
results).~\cite{Adloff:1999nr,Breitweg:2001nz}
Different explanations can be found for this excess ranging from light
bottom squarks and gluinos \cite{Berger:2001mp} to resummed calculations
in the massless evolution scheme \cite{Binnewies:1998vm} and unintegrated
gluon densities.~\cite{Ball:2001pq}

\section{Summary}

In summary, perturbative and non-perturbative QCD continues to be a 
challenge in hard interactions, jet and heavy flavor production.
In many cases it is the theoretical error which is now dominating and requires
a substantial effort in multi-particle NLO, NNLO, and resummed calculations.
The most challenging discrepancies consist in the bottom cross section and
in quarkonium production, where the universality of NRQCD and its color-octet
 mechanism  are seriously challenged from the persisting TEVATRON-HERA
anomaly.

After the luminosity upgrade HERA II will be the only European collider
producing new data on high $Q^2$ and $E_T^2$ jet production, 
bottom differential cross sections and decays, and it will have an increased
discovery reach for physics beyond the Standard Model. At higher energies,
jet and heavy flavor production could be studied at a future THERA
collider.~\cite{Klasen:2001jd}

\section*{Acknowledgments}
The author thanks R.\ Nania for the kind invitation and financial support,
B.A.\ Kniehl, G.\ Kramer, L.\ Mihaila, and M.\ Steinhauser for their
collaboration, and M.\ Dasgupta, M.\ Fontannaz, G.\ Grindhammer, P.\ Hodgson,
J.\ Meyer, J.\ Repond, J.\ Whitmore, M.\ Wing, and Z.\ Trocsanyi for helpful
discussions. Financial support by the Deutsche Forschungsgemeinschaft through
Grant No.\ KL~1266/1-1, by the Bundesministerium f\"ur Bildung und Forschung
through Grant No.\ 05~HT9GUA~3, and by the European Commission through the
Research Training Network {\it Quantum Chromodynamics and the Deep Structure
of Elementary Particles} under Contract No.\ ERBFMRX-CT98-0194 is gratefully
acknowledged.

\end{document}